\documentclass[journal]{IEEEtran}
\usepackage{cite}
\usepackage{ifpdf}
\usepackage{array}
\ifpdf
\usepackage{graphicx}
\usepackage{subfigure}
\usepackage[svgnames]{xcolor}
\else
\usepackage[dvipdfmx]{graphicx}
\usepackage[dvipdfmx,svgnames]{xcolor}
\fi
\usepackage{amsmath,amssymb,amscd}
\usepackage{verbatim}
\usepackage{enumerate}

\newtheorem{prop}{Proposition}
\usepackage{algorithm}
\usepackage{algpseudocode}
\usepackage{tikz}
\begin{document}
\title{On the Trade-Off between Computational Load and Reliability for Network Function Virtualization}

\author{\large Jinkyu~Kang, Osvaldo~Simeone,~\IEEEmembership{Fellow,~IEEE,} and Joonhyuk~Kang,~\IEEEmembership{Member,~IEEE.}
\vspace{-0.5cm}

\thanks{The work of O. Simeone was partially supported by the U.S. NSF through grant CCF-1525629. This research was supported by Basic Science Research Program through the National Research Foundation of Korea(NRF) funded by the Ministry of Science, ICT $\&$ Future Planning(2017R1A2B2012698).

Jinkyu Kang is with the School of Engineering and Applied Sciences (SEAS), Harvard University, Cambridge, MA 02138, USA (Email: jkkang@g.harvard.edu).

O. Simeone is with the Center for Wireless Information Processing (CWIP), ECE Department, New Jersey Institute of Technology (NJIT), Newark, NJ 07102, USA (Email: osvaldo.simeone@njit.edu). 

Joonhyuk Kang is with the Department of Electrical Engineering, Korea Advanced Institute of Science and Technology (KAIST) Daejeon, South Korea (Email: jhkang@ee.kaist.ac.kr).
}
}
\maketitle
\begin{abstract}
Network Function Virtualization (NFV) enables the ``softwarization'' of network functions, which are implemented on virtual machines hosted on Commercial off-the-shelf (COTS) servers. Both the composition of the virtual network functions (VNFs) into a forwarding graph (FG) at the logical layer and the embedding of the FG on the servers need to take into account the less-than-carrier-grade reliability of COTS components. This work investigates the trade-off between end-to-end reliability and computational load per server via the joint design of VNF chain composition (CC) and FG embedding (FGE) under the assumption of a bipartite FG that consists of controller and regular VNFs. Evaluating the reliability criterion within a probabilistic model, analytical insights are first provided for a simplified disconnected FG. Then, a block coordinate descent method based on mixed-integer linear programming is proposed to tackle the joint optimization of CC and FGE. Via simulation results, it is observed that a joint design of CC and FGE leads to substantial performance gains compared to separate optimization approaches.
\end{abstract}

\vspace{-0.2cm}
\section{Introduction} \label{Sec:Intro}
Promoted by the European Telecommunications Standards Institute (ETSI), Network Function Virtualization (NFV) is currently seen as an essential technology to reduce the operators' cost, as well as to provide flexible network deployments. Unlike conventional implementations in which network functions are hard-wired into dedicated hardware, NFV enables the ``softwarization'' of network functions, which are executed on commercial off-the-shelf (COTS) network elements, such as servers and switches. 

The deployment of NFV poses new fundamental challenges that pertain to the following design tasks \cite{Herrera16TNSM}: ({\it{i}}) {\it{VNF Chain Composition}} (CC): Connect the virtual network functions (VNFs) within a forwarding graph (FG) that describes the functional dependence of the VNFs; ({\it{ii}}) {\it{VNF Forwarding Graph Embedding}} (FGE): Instantiate the VNFs on the available network elements; and ({\it{iii}}) {\it{VNFs Scheduling}}: Schedule the execution of the VNFs. 

COTS elements are characterized by a reliability that is significantly lower than the five-nines reliability of carrier-grade equipment, as their operation may be affected by surges in computing load, hardware malfunctions or malicious attacks \cite{ETSI2016, LIU2016WC, Mijumbi2016CST, KHAN15CNSM, Guo14JLT, Qu16CNSM, Yang16RNDM, Casazza17arXiv}. To ensure a desired level of end-to-end (e2e) reliability, previous works  \cite{KHAN15CNSM, Guo14JLT, Casazza17arXiv, Yang16RNDM, Qu16CNSM} allowed the replication of VNFs across multiple servers. In particular, the FGE problem was investigated in \cite{KHAN15CNSM, Guo14JLT} under worst-case e2e reliability guarantees. In contrast, reference \cite{Qu16CNSM} addressed the FGE problem for a general FG and physical layer under a simplified probabilistic framework in which a server cannot be assigned more than one VNF of the given service and the failures of different VNFs are independent. A similar model was also adopted in \cite{Yang16RNDM, Casazza17arXiv}, with \cite{Yang16RNDM} considering a communication delay between two virtual machines.

In this work, we investigate the trade-off between the e2e reliability of a network service, as measured by the probability that the service is correctly executed, and the computational load of the servers. This trade-off arises from the fact that replicating VNFs on servers enhances e2e reliability but at the cost of increasing the computational load. 

The main contributions of this letter are summarized as follows: ({\it{i}}) We formulate the joint CC and FGE optimization problem for bipartite FGs (B-FG) \cite{LIU2016WC} and a generic physical layer within a probabilistic model of reliability that, unlike \cite{Qu16CNSM}, enables the same server to execute multiple VNFs, both controller and regular (Sec. II); ({\it{ii}}) We provide analytical insights into the trade-off between reliability and per-server computational load for a simplified disconnected FG (D-FG) (Sec. \ref{SEC:DisconnectedFG}); ({\it{iii}}) We propose an algorithm based on block coordinate descent (BCD) and mixed-integer linear programming (MILP) to obtain a heuristic solution to the problem of joint CC and FGE optimization (Sec. \ref{SEC:Bipartite FG}); and ({\it{iv}}) Numerical results are presented (Sec. V) gains in terms of e2e unreliability as compared to a separate design.
\begin{figure}[t]
\centering
\includegraphics[width=6.6cm]{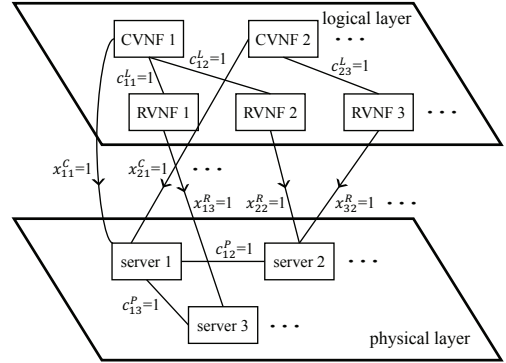}
\vspace{-0.15cm}
\caption{Illustration of an NFV architecture with a bipartite forwarding graph and an arbitrary physical network topology.}
\label{Fig:SystemModel_wDFG}
\vspace{-0.45cm}
\end{figure}

\vspace{-0.2cm}
\section{System Model} \label{Sec:SM}
We consider an NFV-based application in which the physical layer is composed of $N_S$ physical elements, which will be referred to as servers as in \cite{LIU2016WC}, and the logical layer of the given network service is composed of $N_V$ VNFs, as shown in Fig. \ref{Fig:SystemModel_wDFG}. We denote the set of all servers and VNFs as $\mathcal{N}_S = \{1, \dots, N_S\}$ and $\mathcal{N}_V = \{1, \dots, N_V\}$, respectively. The overprovisioning rate $r=N_S/N_V$ is the ratio between the number of servers and VNFs. Assuming that the servers are strategically distributed \cite{ETSI2016, LIU2016WC}, we postulate that each server $s \in \mathcal{N}_S$ fails independently with probability $p_s$ \cite{Qu16CNSM, Yang16RNDM}. This can be related to the mean time between failures as discussed in \cite{ETSI2016}.

At the logical layer, the VNFs are connected by means of an FG that describes the functional relationship among VNFs. We focus on the B-FG considered in \cite{LIU2016WC}, in which each regular VNF (RVNF) is managed by a controller VNF (CVNF). We define as $\mathcal{N}_{V, R}$ the set of the $N_{V, R}$ RVNFs and as $\mathcal{N}_{V, C}$ the set of the $N_{V, C} \le N_{V, R}$ CVNFs with ${N}_V = {N}_{V, C} + {N}_{V, R}$. The FG, which is to be optimized, is defined via the following binary variables: 
\begin{equation}
\hspace{0.05cm} c^L_{uv} \hspace{-0.1cm} = \hspace{-0.05cm} {\bf{1}} \hspace{-0.05cm} \left ( \text{CVNF $u \hspace{-0.1cm}\in\hspace{-0.1cm} \mathcal{N}_{V, C}$ is connected to RVNF $v \hspace{-0.1cm}\in\hspace{-0.1cm} \mathcal{N}_{V, R}$} \right ) \hspace{-0.05cm}.\hspace{-0.3cm}
\end{equation}
Each CVNF $u \in \mathcal{N}_{V, C}$ can manage at most $C_u$ RVNFs, and hence we have the constraint $\sum_{v \in \mathcal{N}_{V, R}} c^L_{uv} \le C_u$. Note that the condition $\sum_{u \in \mathcal{N}_{V, C}} C_u \ge N_{V, R}$ is required. At the physical layer, the connectivity between servers, which is fixed, is defined by the following binary variables:
\begin{equation} \label{BinaryVar:phylayer}
c^P_{st} = {\bf{1}} \left( \text{server $s$ is connected to server $t$} \right ).
\end{equation}

The virtualization layer maps each VNF into one or more servers. We denote as $x_{vs}$ the binary variable that indicates whether VNF $v \in \mathcal{N}_{V,R} \cup \mathcal{N}_{V, C}$ is instantiated on server $s$, namely
\begin{equation} \label{BinaryVar:virtuallayer}
x_{vs} = {\bf{1}} \left( \text{VNF $v$ is instantiated on server $s$} \right),
\end{equation}
which can be divided into $x_{us}^C$ for CVNF $u \in \mathcal{N}_{V, C}$ and $x_{ws}^R$ for RVNF $w \in \mathcal{N}_{V, R}$. Note that, in order to enhance the e2e reliability, we allow VNFs to be replicated across different servers \cite{Casazza17arXiv}. Mapping (\ref{BinaryVar:virtuallayer}) must ensure that, if the edge CVNF $u$ $-$ RVNF $v$ exists in the FG and if CVNF $u \in \mathcal{N}_{V, C}$ is instantiated on server $s$ while RVNF $v \in \mathcal{N}_{V, R}$ is instantiated on server $t$, then there must exist a physical link $s - t$ in order for the logical link to be active \cite{LIU2016WC}. Mathematically, we have the constraint $c^P_{st}=1$ if $x_{us}^C x_{vt}^R c^{L}_{uv}=1$, which can be reformulated as $x_{us}^C x_{vt}^R c^L_{uv} = 0$ if $c^P_{st} = 0$ for any $u \in \mathcal{N}_{V, C}$, $v \in \mathcal{N}_{V, R}$, and $s, t \in \mathcal{N}_S$.

A successful completion of the network service occurs if all VNFs are correctly executed on at least one server. Hence, the probability of successful completion of the application, or e2e reliability, can be defined as 
\begin{equation} \label{Prob_Succ}
P_S = \sum_{{\bf{f}} \in \{0, 1\}^{N_S}} \text{P}({\bf{f}}) \prod_{v=1}^{N_V} Q_v({\bf{f}}, {\bf{c}}^L, {\bf{x}}),
\end{equation}
where the vector ${\bf{f}} = [f_1, \dots, f_{N_S}]$ collects the indicators $f_s$, which are defined as $f_s = {\bf{1}} \left( \text{server $s$ is on} \right)$ and hence we have the probability distribution $\text{P}({\bf{f}})= \prod_{s \in \mathcal{S}_0} p_s \prod_{t \in \mathcal{S}_1} (1- p_t)$ with $\mathcal{S}_0 = \{s|f_s=0\}$ and $\mathcal{S}_1 = \{s|f_s=1\}$; and $Q_v({\bf{f}}, {\bf{c}}^L, {\bf{x}})$ indicates whether the RVNF $v$ is successfully executed as a function of ${\bf{f}}$, ${\bf{c}}^L = \{c^L_{vu}\}_{v \in \mathcal{N}_{V, C}, u \in \mathcal{N}_{V, R}}$, and ${\bf{x}} = \{x_{vs}\}_{v \in \mathcal{N}_V, s \in \mathcal{N}_S}$, that is, we have $Q_v({\bf{f}}, {\bf{c}}^L, {\bf{x}}) = {\bf{1}} \left ( \text{VNF $v$ is executed correctly} \right)$. This quantity can be expressed as 
\begin{equation} \label{Def_IndicatorVNF:regVNF}
Q_v ({\bf{f}}\hspace{-0.05cm},\hspace{-0.05cm} {\bf{c}}^L\hspace{-0.05cm},\hspace{-0.05cm} {\bf{x}}) \hspace{-0.05cm} =\hspace{-0.05cm} \left \lVert \sum_{s, t \in \mathcal{N}_S} \hspace{-0.07cm} \sum_{u \in \mathcal{N}_{V, C}} \hspace{-0.2cm} f_s x_{us}^C  c^L_{uv} c^P_{st} f_t x_{vt}^R \right \rVert_0 \hspace{-0.15cm} = \hspace{-0.05cm} \left \lVert \sum_{u \in \mathcal{N}_{V, C}} \hspace{-0.2cm} \psi_{uv} \right \rVert_0
\end{equation}
where $\psi_{uv} = \sum_{s, t \in \mathcal{N}_S} f_s x_{us}^C  c^L_{uv} c^P_{st} f_t x_{vt}^R= c^L_{uv} ( {\bf{f}} \circ {\bf{x}}_u^C )^T {\bf{C}}^P ( {\bf{f}} \circ {\bf{x}}_v^R )$; ${\bf{x}}_u^C = [x_{u 1}^C, \dots, x_{u N_S}^C]^T$; ${\bf{x}}_v^R = [x_{v 1}^R, \dots, x_{v N_S}^R]^T$; ${\bf{C}}^P$ is a $N_S \times N_S$ matrix with the elements $[{\bf{C}}^P]_{s, t} = c^P_{st}$; and $\circ$ denotes the Hadamard (element-wise) product, e.g., ${\bf{f}} \circ {\bf{x}}_v^R = [f_1 x_{v1}^R, \dots,$ $f_{N_S} x_{v N_{S}}^R]^T$. Equation (\ref{Def_IndicatorVNF:regVNF}) follows since a RVNF $v$ is correctly executed if the CVNF $u$ supporting the RVNF $v$ is instantiated in an active server, which is connected to at least one of the active server running the RVNF $v$.

We focus on the problem of joint CC and FGE, which amounts to the joint optimization of the logical link connectivity ${\bf{c}}^L$ (CC) and the optimal virtualization mapping ${\bf{x}}$ (FGE). We are specifically interested in maximizing the e2e reliability under a maximum computational load constraint $L$ given the physical network topology ${\bf{c}}^P = \{ c^P_{st} \}_{s, t \in \mathcal{N}_S}$ and the CVNF capacity limit $C_u$. This optimization problem is formulated as 
\begin{subequations} \label{OP}
\begin{eqnarray}
\hspace{-0.8cm} \label{OP:ob} \underset { {\bf{x}}, {\bf{c}}^L }{\textrm{maximize}} && \hspace{-0.8cm}\sum_{{\bf{f}} \in \{0, 1\}^{N_S}} \hspace{-0.3cm} \text{P}({\bf{f}}) \hspace{-0.07cm} \cdot \hspace{-0.07cm} {\bf{1}} \hspace{-0.1cm} \left ( \hspace{-0.1cm} N_{V, R} \hspace{-0.05cm}-\hspace{-0.05cm} \left \lVert \begin{subarray}{c}  \sum_{u \in \mathcal{N}_{V, C}} \psi_{u1} \\  \vdots \\ \sum_{u \in \mathcal{N}_{V, C}} \psi_{u N_{V, R}}  \end{subarray}  \right \rVert_0 \le 0 \right)\hspace{-0.1cm} \\
\hspace{-0.8cm} \label{OP:latencyConst} {\textrm{s.t.}} \hspace{0.5cm} && \hspace{-0.5cm} \sum_{v=1}^{N_V} x_{vs} \le L, \,\,\,\,\,\,\,\,\,\,\,\,\,\,\,\,\,\,\,\,\,  \text{for} \,\,\, s \in \mathcal{N}_S \\
\hspace{-0.8cm} \label{OP:binaryConst_x} && \hspace{-0.5cm} x_{vs} \in \{0, 1\}, \,\,\,\,\,\,\,\,\,\,\,\,\,\,\,\,\,\,\,\,\, \text{for} \,\,\, v \in \mathcal{N}_V, \,\, s \in \mathcal{N}_S \\
\hspace{-0.8cm} \label{OP:Log_Topology} && \hspace{-0.5cm} c_{uv}^L \in \{0, 1\}, \sum_{u \in \mathcal{N}_{V, C}} c^L_{uv} = 1, \sum_{v \in \mathcal{N}_{V, R}} c^L_{uv} \le C_u, \\
\hspace{-0.8cm} \label{OP:Phy_Topology} && \hspace{-0.5cm} x_{us}^C x_{vt}^R c^{L}_{uv}=0, \,\, \text{for} \,\, (s, t) \in \{ (s, t) | c^P_{st}=0\},
\end{eqnarray}
\end{subequations}
where (\ref{OP:Log_Topology})-(\ref{OP:Phy_Topology}) apply to $\forall u \in \mathcal{N}_{V, C}$, $\forall v \in \mathcal{N}_{V, R}$, and $\forall s \in \mathcal{N}_S$, and it can be verified that (\ref{OP:ob}) equals (\ref{Prob_Succ}). 

\emph{Notation}: $\Vert {\bf{x}} \Vert_0 = 0$ if ${\bf{x}}=0$ and $\Vert {\bf{x}} \Vert_0 = 1$ otherwise. ${\bf{1}}(x) = 1$ if $x$ is true and ${\bf{1}} (x) = 0$ otherwise.
\section{Disconnected Forwarding Graph Embedding} \label{SEC:DisconnectedFG}
In this section, in order to obtain initial insights into the FGE problem, we consider a simplified scenario with a D-FG. This can be seen a special case of problem (\ref{OP}) in which all VNFs can be equivalently considered as RVNFs with the role of controller, i.e., $\mathcal{N}_{V, C} = \emptyset$, the constraints (\ref{OP:Log_Topology})-(\ref{OP:Phy_Topology}) are not present, and $p_s = p$ for $\forall s \in \mathcal{N}_S$. Note that CC is not applicable to a D-FG since the FG is fixed. For FGE problem, we derive a lower bound on the optimal solution of problem (\ref{OP}) via a union bound argument. To elaborate, we observe that a VNF $v$ is correctly executed if at least one server to which it is allocated is on, and hence the probability of this event is $P_{S, v} = p^{\sum_{s=1}^{N_S} x_{vs}}$. Accordingly, the e2e reliability $P_S$ in (\ref{Prob_Succ}) can be lower bounded by the union bound as $P_S \ge 1 - \sum_{v=1}^{N_V} p^{\sum_{s=1}^{N_S} x_{vs}}$. With this lower bound in lieu of the exact probability $P_S$ in (\ref{Prob_Succ}), the problem can be reformulated as 
\begin{subequations} \label{OP_DFG_OV}
\begin{eqnarray}
\underset { {\bf{x}} }{\textrm{minimize}} && \sum_{v=1}^{N_V} p^{\sum_{s=1}^{N_S} x_{vs}} \\
{\textrm{s.t.}} \hspace{0.5cm} && (\ref{OP:latencyConst})-(\ref{OP:binaryConst_x}).
\end{eqnarray}
\end{subequations}
The solution of this problem identifies a trade-off between the e2e reliability and the computational load $L$ as a function of the overprovisioning rate $r$ as described in the following proposition.
\prop \label{Prop1} The optimal solution of problem (\ref{OP_DFG_OV}) yields the following lower bound on the e2e reliability obtained from (\ref{OP}) for a D-FG
\begin{equation} \label{Prop1:eq}
P_S \ge 1 - N_V p^{r \cdot \min (L, N_V)},
\end{equation} 
for all values of overprovisioning rate $r$ and computational load $L$ such that $r L$ is an integer.
\begin{IEEEproof}
We prove by contradiction that any solution with $\sum_{v=1}^{N_V} x_{vs} =  L$ for all $s \in \mathcal{N}_S$ and $\sum_{s=1}^{N_S} x_{vs} = r L$ for all $v \in \mathcal{N}_V$, is optimal if $r L $ is an integer. First, it is easily proved that the latency constraint (\ref{OP:latencyConst}) must be satisfied with equality by any optimal solution and hence the first condition must hold. For the latter condition, assume that an optimal mapping satisfies $\sum_{s=1}^{N_S} x_{vs} < r  L$ for some $v \in \mathcal{N}_V$. Since we have $\sum_{v=1}^{N_V} x_{vs} = L$ for all $s = 1, \dots, N_S$, this condition can be expressed as 
\begin{equation} \label{Contradiction_cond}
\sum_{v = 1}^{N_V} p^{\sum_{s=1}^{N_S} x_{vs}} < N_V p^{r L},
\end{equation}
where $\sum_{s=1}^{N_S} x_{vs}$ is the number of server in which VNF $v$ is instantiated. However, by Jensen's inequality and the fact that an exponential function $p^z$ with $0 \le p \le 1$ is convex in $z \ge 0$, we obtain the inequality $\sum_{v = 1}^{N_V} p^{\sum_{s=1}^{N_S} x_{vs}}/N_V \ge p^{r L}$, which implies that (\ref{Contradiction_cond}) cannot be satisfied, hence concluding the proof.
\end{IEEEproof}
The proposition above suggests that the e2e unreliability $1-P_S$ decreases exponentially as $p^{rL}$ for increasing overprovisioning rate $r$ and computational load $L$. We will verify this insight in Sec. \ref{Sec:Simulation} via numerical results. 


\section{Bipartite Forwarding Graph \\ Composition and Embedding} \label{SEC:Bipartite FG}
In this section, we consider the general CC and FGE problem (\ref{OP}). To this end, we make two approximations. First, we substitute the non-convex indicator function in the objective function with the concave lower bound given by the hinge loss function as ${\bf{1}}(y \le 0 ) \ge \min (1, 1 - y)$. Second, we approximate the $l_0$ norm with the $l_1$ norm, as it is often done in related problems. With these approximations, the problem (\ref{OP}) can be written using the epigraph formulation as 
\begin{subequations} \label{OP_BFG2}
\begin{eqnarray}
\label{OP_BFG2:ob} \underset {{\bf{x}}, {\bf{c}}^L, {\bf{t}} }{\textrm{maximize}} && \hspace{-.5cm} \sum_{{\bf{f}} \in \{0, 1\}^{N_S}} \text{P}({\bf{f}}) \cdot t _{\bf{f}} \\
\label{OP_BFG:EpiConst1} {\textrm{s.t.}} \hspace{0.5cm} && \hspace{-.5cm} t _{\bf{f}} \le 1, \\
\label{OP_BFG:EpiConst2} && \hspace{-.5cm}  t _{\bf{f}} \le 1- N_{V, R} + \sum_{u \in \mathcal{N}_{V, C}} \sum_{v \in \mathcal{N}_{V, R}} \psi_{uv}, \\
&& \hspace{-.5cm} (\ref{OP:latencyConst})-(\ref{OP:Phy_Topology}).
\end{eqnarray}
\end{subequations}
In (\ref{OP_BFG2}), we have defined a vector of auxiliary variables ${\bf{t}} = \{t_{\bf{f}} \}_{{\bf{f}} \in \{0, 1\}^{N_S}}$ and we used the equality as $\lVert [ \sum_{u \in \mathcal{N}_{V, C}} c^L_{u1} ( {\bf{f}} \circ {\bf{x}}_u^C )^T {\bf{C}}^P ( {\bf{f}} \circ {\bf{x}}_1^R ), \cdots,\sum_{u \in \mathcal{N}_{V, C}} c^L_{u N_{V, R}} ( {\bf{f}} \circ {\bf{x}}_u^C )^T {\bf{C}}^P ( {\bf{f}} \circ {\bf{x}}_{N_{V, R}}^R ) ]^T \rVert_1 = \sum_{u \in \mathcal{N}_{V, C}}$ $\sum_{v \in \mathcal{N}_{V, R}} c^L_{uv} ( {\bf{f}} \circ {\bf{x}}_u^C )^T {\bf{C}}^P ( {\bf{f}} \circ {\bf{x}}_v^R )$. To address problem (\ref{OP_BFG2}), we approach the joint optimization via a BCD method, whereby at each iteration we optimize iteratively over variables $({\bf{x}}^R, {\bf{c}}^L)$ and variables $({\bf{x}}^C, {\bf{c}}^L)$ by solving a MILP. The overall algorithm is summarized in Algorithm \ref{al1}. We now discuss how to perform the inner optimization over $({\bf{x}}^R, {\bf{c}}^L)$, and then cover also in a similar way the optimization over $({\bf{x}}^C, {\bf{c}}^L)$.

For the optimization over $({\bf{x}}^R, {\bf{c}}^L)$, we follow the approach in \cite{Qu16CNSM} of defining the auxiliary binary variables $z^R_{uvs} = x^R_{vs} c^L_{uv}$ for all $u \in \mathcal{N}_{V, C}$, $v \in \mathcal{N}_{V, R}$, $s \in \mathcal{N}_S$. With this definition, the non-linear constraints (\ref{OP_BFG:EpiConst2}), (\ref{OP:Phy_Topology}) become linear in ${\bf{z}}^R = \{z^R_{uvs}\}$ as
\begin{subequations} \label{OP_BFG2:RevisedConst}
\begin{equation}
t _{\bf{f}} \hspace{-0.05cm} \le \hspace{-0.05cm} 1- N_{V, R} + \hspace{-0.5cm} \sum_{v \in \mathcal{N}_{V, R}, u \in \mathcal{N}_{V, C}} \hspace{-0.5cm} ( {\bf{f}} \circ {\bf{x}}^C_u )^T {\bf{C}}^P ( {\bf{f}} \circ {\bf{z}}_{uv}^R ), \,\, \text{for} \,\,  {\bf{f}} \in \{0, 1\}^{N_S},
\end{equation}
\begin{equation}
\text{and} \,\,\,\,\,\,\, x_{ut}^C z^R_{uvs} =0, \,\,\,\, \text{for} \,\,\,\, (s, t) \in \{ (s, t) | c^P_{st}=0\},
\end{equation}
\end{subequations}
respectively, for $u \in \mathcal{N}_{V, C}$ and $v \in \mathcal{N}_{V, R}$, with ${\bf{z}}_{uv}^R = \{z^R_{uvs}\}_{s \in \mathcal{N}_S}$. Moreover, the variables $z^R_{uvs}$ are characterized by the conditions
\begin{equation}\label{OP_BFG2:AdditionalConst}
z^R_{uvs} \le x^R_{vs}, \,\,\, z^R_{uvs} \le c^L_{uv}, \,\,\, \text{and} \,\,\, z^R_{uvs} \ge x^R_{vs} + c^L_{uv} - 1,
\end{equation}
for $u \in \mathcal{N}_{V, C}$, $v \in \mathcal{N}_{V, R}$, and $s \in \mathcal{N}_S$. The resulting problem is given as
\begin{subequations} \label{OP_BFG3}
\begin{eqnarray}
\underset {{\bf{x}}^R, {\bf{c}}^L, {\bf{z}}^R, {\bf{t}} }{\textrm{maximize}} && \sum_{{\bf{f}} \in \{0, 1\}^{N_S}} \text{P}({\bf{f}}) \cdot t _{\bf{f}} \\
\label{OP_BFG2:Const} {\textrm{s.t.}} \hspace{0.5cm} && (\ref{OP:latencyConst}) - (\ref{OP:Log_Topology}), (\ref{OP_BFG:EpiConst1}), (\ref{OP_BFG2:RevisedConst}), (\ref{OP_BFG2:AdditionalConst}).
\end{eqnarray}
\end{subequations}
Problem (\ref{OP_BFG3}) is a MILP which can be solved using efficient numerical tools. Analogously, for the latter inner problem over $( {\bf{x}}^C$, ${\bf{c}}^L )$, the problem can be reformulated as the MILP (\ref{OP_BFG3}) upon switching the superscripts ``$R$'' and ``$C$'' with ${\bf{z}}_{uv}^C = \{z^C_{uvs}\}_{s \in \mathcal{N}_S}$, for $u \in \mathcal{N}_{V, C}$ and $v \in \mathcal{N}_{V, R}$. 

\begin{algorithm} [h!]
\begin{algorithmic}
\caption{Joint CC-FGE optimization} \label{al1}
\State {\textbf{Initialization}}: Initialize ${\bf{x}}^{R, (0)} = {\bf{x}}^{C, (0)} = {\bf{0}}$ and set $n=0$.
\State {\textbf{Repeat}}
\State \indent $n \gets n+1$
\State \indent Compute ${\bf{x}}^{R, (n)}$, ${\bf{c}}^L$, ${\bf{z}}^{R}$ and ${\bf{t}}$ by solving the problem (\ref{OP_BFG3}) with ${\bf{x}}^C \gets {\bf{x}}^{C, (n-1)}$.
\State \indent Compute ${\bf{x}}^{C, (n)}$, ${\bf{c}}^L$, ${\bf{z}}^{C}$ and ${\bf{t}}$ by solving the problem reformulated from (\ref{OP_BFG3}) with ${\bf{x}}^R \gets {\bf{x}}^{R, (n)}$.
\State {\textbf{Until}} a convergence criterion is satisfied. 
\State {\textbf{Solution}}: ${\bf{x}}^C \gets {\bf{x}}^{C, (n)}$, ${\bf{x}}^R \gets {\bf{x}}^{R, (n)}$ and ${\bf{c}}^L$.
\end{algorithmic}
\end{algorithm}

\vspace{-0.5cm} 
\section{Simulation Results} \label{Sec:Simulation}
In this section, we present numerical results that encompass both the simplified D-FG considered in Sec. \ref{SEC:DisconnectedFG} and the more general B-FG investigated in Sec. \ref{SEC:Bipartite FG}. For the D-FG case, we consider the union bound (\ref{Prop1:eq}) and the solution obtained for the FGE problem via Algorithm \ref{al1} (setting $N_V=N_{V,R}$, ${\bf{x}} = {\bf{x}}^C = {\bf{x}}^R$ and ${\bf{c}}^L=1$, and removing the constraints (\ref{OP:Log_Topology})-(\ref{OP:Phy_Topology})). For the B-FG case, we compare the performance of the joint CC and FGE optimization in Algorithm \ref{al1} with FGE optimization only. For the latter case, the fixed FG is composed in either of these two ways: ({\it{i}}) CCmin: Each CVNF monitors the minimum number of RVNFs that is compatible with the requirement that each RVNF is assigned a CVNF; and ({\it{ii}}) CCmax: Each CVNF $u$ monitors the maximum number of RVNFs allowed by the capacity limit $C_u$ and by the mentioned requirement on RVNF-CVNF assignment. Note that an FG built according to CCmin provides more flexibility for FGE optimization in sparsely connected physical layers with a sufficiently large number of servers, since each CVNF only monitors a small number of RVNFs. Instead, CCmax allows larger diversity gains to be obtained in the presence of a well-connected physical layer, since a smaller number of CVNFs can be replicated more aggressively on the servers. Throughout, we consider independently generated physical connectivity variables (\ref{BinaryVar:phylayer}) with $\textrm{Pr} (c^P_{st}=1) = p^P$ for all server pairs $s, t \in \mathcal{N}_S$ and $s \neq t$, and the performance is averaged over these variables. Moreover, every server has the same failure probability, i.e., $p_s = p$ for all $s \in \mathcal{N}_S$ and each CVNF $u \in \mathcal{N}_{V, C}$ has the same capacity $C_u = C$.

\begin{figure}[t!]
\centering
\vspace{-0.0cm}
\includegraphics[width=7.4cm]{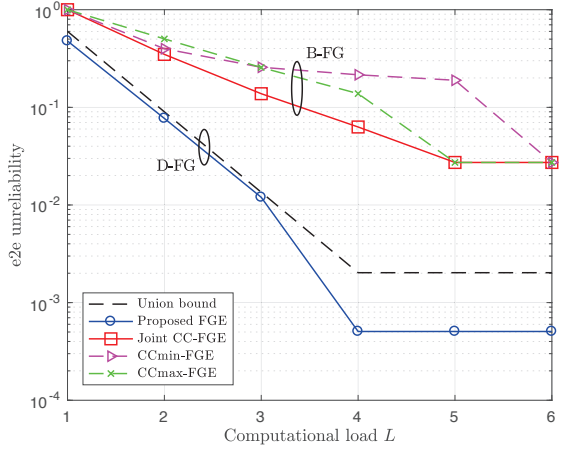}
\vspace{-0.3cm}
\caption{Outage probability vs. computational load $L$ for $p=0.15$, $N_S = 4$, $N_{V,C}=2$, $N_{V,R}=4$, $C=4$, and $p^P = 0.8$.}
\vspace{-0.4cm}
\label{Fig:OutageProb_vsL}
\end{figure}

\begin{figure}[t!]
\centering
\includegraphics[width=7.4cm]{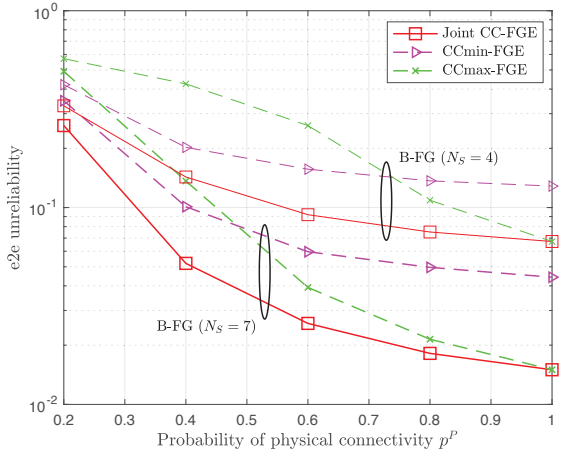}
\vspace{-0.3cm}
\caption{Outage probability vs. probability of physical connectivity $p^P$ for $p=0.05$, $N_{V, C}=2$, $N_{V,R}=4$, $L=3$, and $C=4$.}
\vspace{-0.4cm}
\label{Fig:OutageProb_vsp_p}
\end{figure}


The outage probability is plotted versus the computational load $L$ in Fig. \ref{Fig:OutageProb_vsL} for $p=0.15$, $N_S = 4$, $N_{V,C}=2$, $N_{V,R}=4$, $C=4$, and $p^P = 0.8$. A first observation is that the union bound obtained for D-FG in Proposition \ref{Prop1} well reflects the dependence of the e2e unreliability on the computational load $L$, and that it closely predicts the performance obtained by the proposed scheme. Given the large probability $p^P$, the physical layer is well connected and, as discussed, CCmax-FGE is advantageous compared to CCmin-FGE for the B-FG case, unless the computational load $L$ is too small, due to its larger diversity gain. Moreover, the advantage of joint CC-FGE optimization is especially pronounced as compared to FGE-only optimization in the region of intermediate computational load, owing to the scarcity of computational resources.

In Fig. \ref{Fig:OutageProb_vsp_p}, we show the impact of the probability of the physical connectivity $p^P$ for the B-FG model with $p=0.05$, $N_{V, C}=2$, $N_{V,R}=4$, $L=3$ and $C=4$. As discussed, for low $p^P$ physical connectivity, the CCmin approach outperforms CCmax, due to its enhancement of flexibility. Its e2e reliability is seen to tend to that of joint optimization for $p^P \to 1$. The opposite is instead true for larger $p^P$, as, in this regime, CCmax can use a smaller number of servers, which is particularly advantageous when $N_S$ is small. Finally, joint CC-FGE optimization provides significant gains for intermediate values of $p^P$. 

\vspace{-0.2cm}
\section{Conclusions}
In this letter, we have studied the trade-off between end-to-end reliability and computational load in NFV for disconnected or bipartite FG models. Our main contributions are as follows: ({\it{i}}) the end-to-end unreliability for D-FGs is approximately proportional to $p^{rL}$, where $p$ is the COTS servers' unreliability, $r$ is the overprovisioning ratio, and $L$ is the computational load per server, as long as $L$ is small enough (Proposition \ref{Prop1} and Fig. \ref{Fig:OutageProb_vsL}); and ({\it{ii}}) the joint design of CC and FGE has significant advantages for B-FGs in the regime of intermediate computational load and low physical connectivity (Fig. \ref{Fig:OutageProb_vsL} and Fig. \ref{Fig:OutageProb_vsp_p}). Future work may include extensions that account for other resource limitations, such as memory, and comparisons with experimental results. 


\vspace{-0.2cm}
\bibliographystyle{IEEEtran}
\bibliography{refKJK}

\end{document}